\documentclass[11pt]{article}
\usepackage{moriond,epsfig}

\def\Journal#1#2#3#4{{#1} {\bf #2}, #3 (#4)}

\def\PRL{\em Phys. Rev. Lett.}

\def\be{\begin{equation}}
\def\ee{\end{equation}}
\def\bea{\begin{eqnarray}}
\def\eea{\end{eqnarray}}

\def\fbi{fb$^{-1}$}
\begin{document}
\vspace*{4cm}
\title{SEARCHES FOR NEW PHENOMENA WITH LEPTON FINAL STATES AT THE TEVATRON}

\author{ T. ADAMS \\ {for the CDF and D0 Collaborations}}

\address{Department of Physics, Florida State University, \\
Tallahassee, FL, 32306, USA}

\maketitle\abstracts{
Numerous searches for new phenomena have been carried out using
data from proton-antiproton collisions at Fermilab's Tevatron.
Final states with leptons give signatures which are relatively
unique and generally have small backgrounds.  We present many
of the latest results from the CDF and D0 collaborations from
0.4-1.2 fb$^{-1}$ of data.  Topics include supersymmetry,
extra gauge bosons, Randall-Sundrum gravitons, excited electrons
and neutral, long-lived particles.
}

New phenomena searches with leptons favor analyses with relatively
small or well-understood backgrounds.  At the Tevatron, the CDF
and D0 collaborations have strong programs of searches with 
leptonic final states in a wide range of topics.  Some of the
most recent results are discussed here.  All limits presented
are at the 95\% CL.

\section{Charginos and Neutralinos in Trileptons}

Supersymmetry has been a popular extension to the standard
model for several decades and numerous searches for 
evidence of any of the superpartners have been carried out.
At hadron colliders, a unique signature for supersymmetry 
comes in the trilepton (``three lepton'') final states.  If
the masses lie in the correct region, proton-antiproton
collisions can produce charginos and neutralinos in
association:

\begin{equation}
 p\bar{p} \rightarrow \tilde{\chi}_{1}^\pm \tilde{\chi}_2^0
\end{equation}

\noindent
with decay modes

\begin{equation}
 \tilde{\chi}_{1}^\pm \rightarrow \ell^\pm \nu \tilde{\chi}_1^0  \hspace{1cm}
 \tilde{\chi}_2^0 \rightarrow \ell^\pm \ell^\mp \tilde{\chi}_1^0
\end{equation}

In the standard model, trilepton final states are only produced
by rare processes (such as di-bosons) which means these searches
will naturally have small backgrounds.  The challenge lies in
the inefficiency to uniquely identify all three leptons.  The
solution is to use three search techniques: (1) observe
all three leptons; (2) observe two leptons and a third isolated
track; (3) observe two same-signed leptons.  By combining all
three search methods and combinations of electrons, muons and
isolated tracks, the experiments improve the sensitivity to
discovery.

\begin{figure}
 %\rule{5cm}{0.2mm}\hfill\rule{5cm}{0.2mm}
 %\vskip 2.5cm
 %\rule{5cm}{0.2mm}\hfill\rule{5cm}{0.2mm}
 \unitlength1cm
 \begin{picture}(15.0,7.0)
  \put(0.0,0.2){\psfig{figure=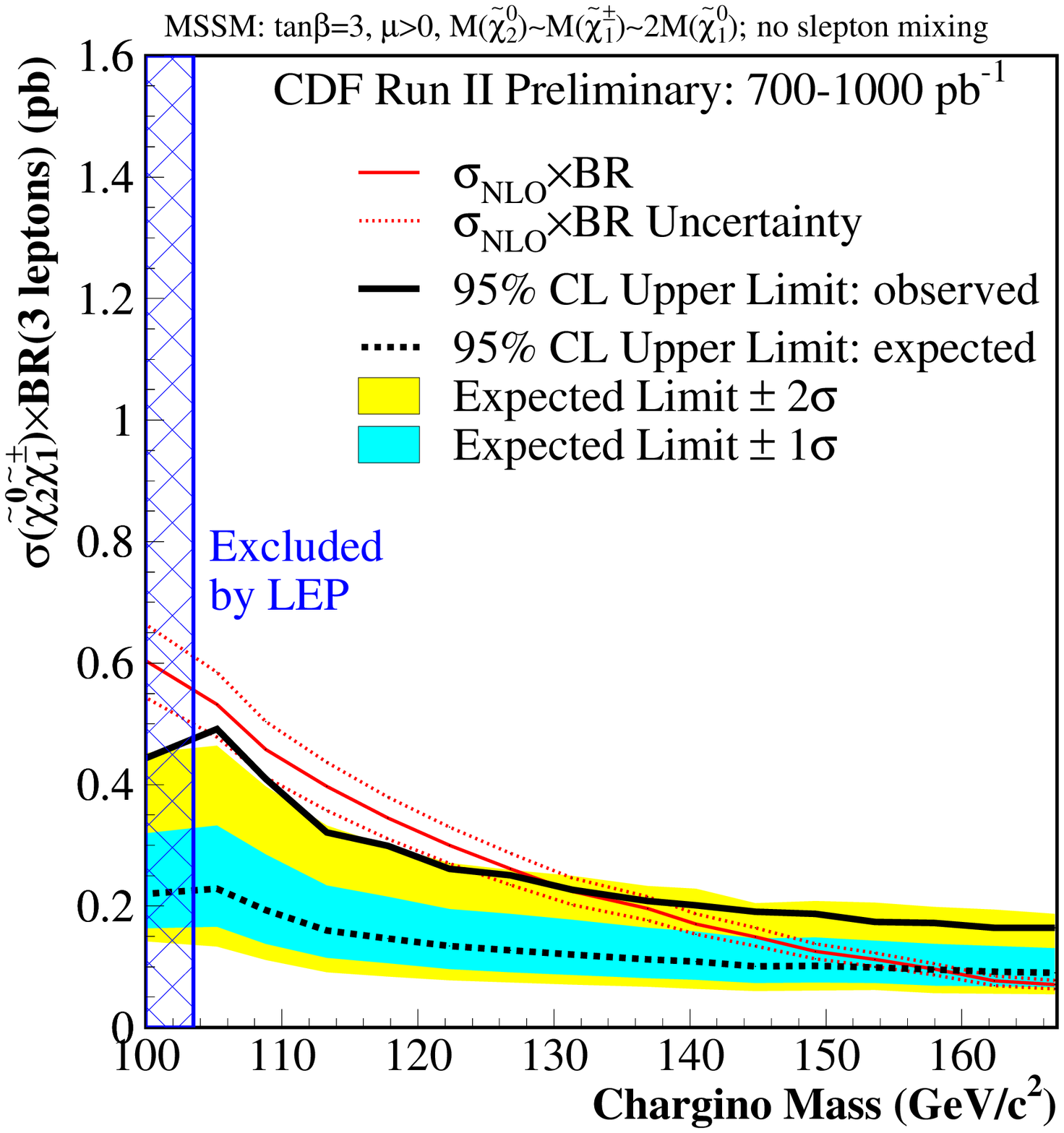,height=6.25cm}}
  \put(6.2,0){\psfig{figure=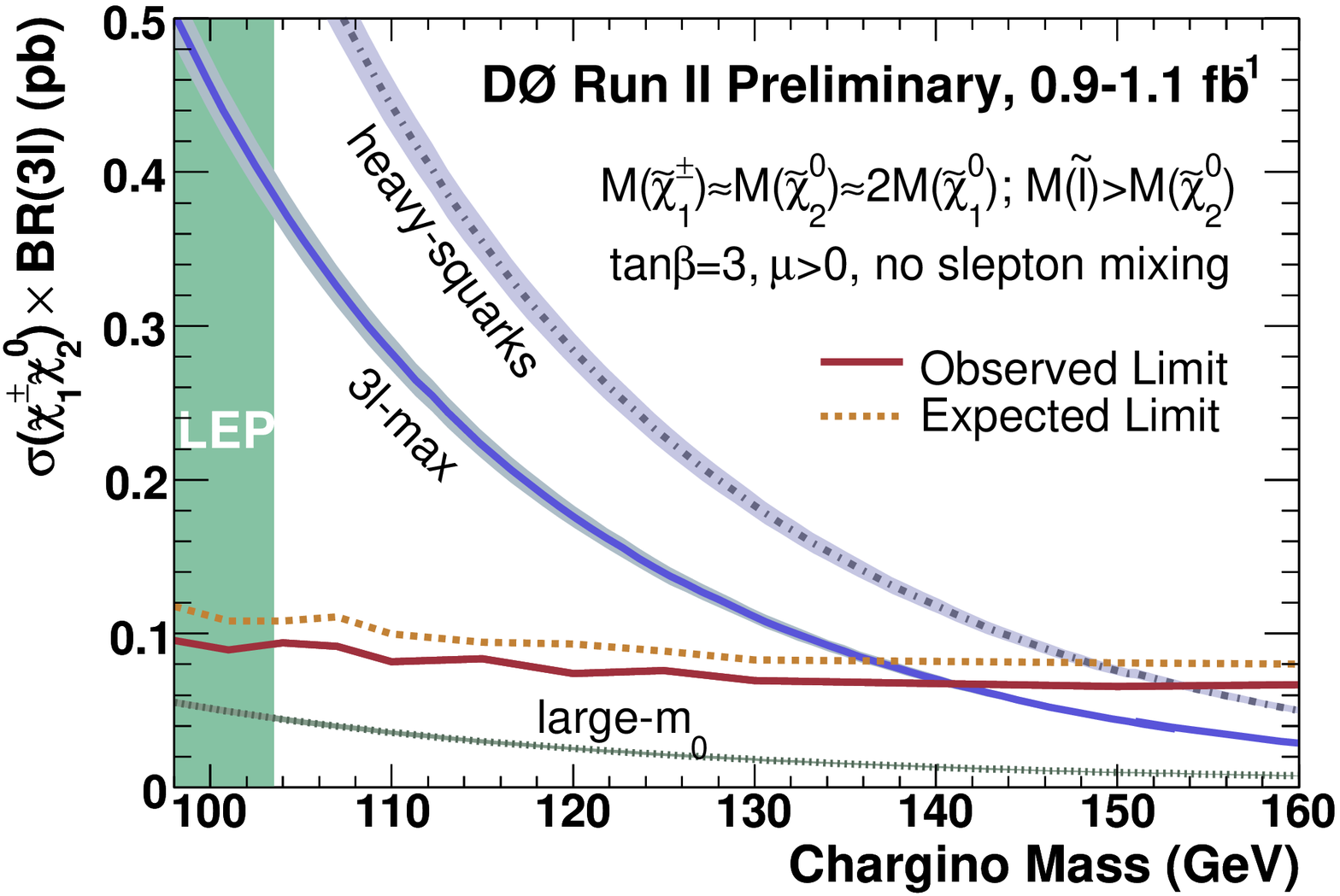,height=6.25cm}}
  \put(4.5,2.5){\Large (a)}
  \put(13.5,2.5){\Large (b)}
 \end{picture}
 \caption{Limits on SUSY production of charginos and neutralinos.  
   (a) CDF limit using a model of MSSM without slepton mixing and
   $m_0 = 60$ GeV;
   (b) D0 limit using three mSUGRA inspired models.
  \label{fig:trileptons}}
\end{figure}

CDF has preformed searches in 14 different channels~\cite{bib:cdftrilepton}  
ranging in luminosity 0.7-1.1 \fbi.
While a slight excess of data
vs. background is observed, there is no strong evidence of
supersymmetry.  Therefore limits on the production 
cross-section times branching ratio can be set.  CDF interprets
this within three different mSUGRA inspired models with
$m_0 = 60$ GeV: (A) mSUGRA;
(B) MSSM without slepton mixing;
(C) MSSM with lepton branching ratio set to same as W/Z.
For model (B) a limit on the $\tilde{\chi}_1^\pm$ mass greater than
130 GeV is set (Fig.~\ref{fig:trileptons}(a)).

D\O\ has preformed four searches~\cite{bib:d0trilepton} 
 with luminosity 1.0-1.1 \fbi \ with no excess of data observed.
Three mSUGRA-inspired models (with no slepton mixing)
are explored: (1) large $m_0$ where W/Z decays dominate;
(2) 3$\ell$-max with slepton mass slightly larger than the
$\tilde{\chi}_2^0$ mass; 
(3) heavy squarks where scalar mass unification is relaxed
(Fig.~\ref{fig:trileptons}(b)).
For the 3$\ell$-max model, a limit of
$M(\tilde{\chi}_1^\pm)$ $>$ 141 GeV is found.

\section{$W^\prime$}

\begin{figure}
 \unitlength1cm
 \begin{picture}(15.0,7.0)
  \put(0.5,0){\psfig{figure=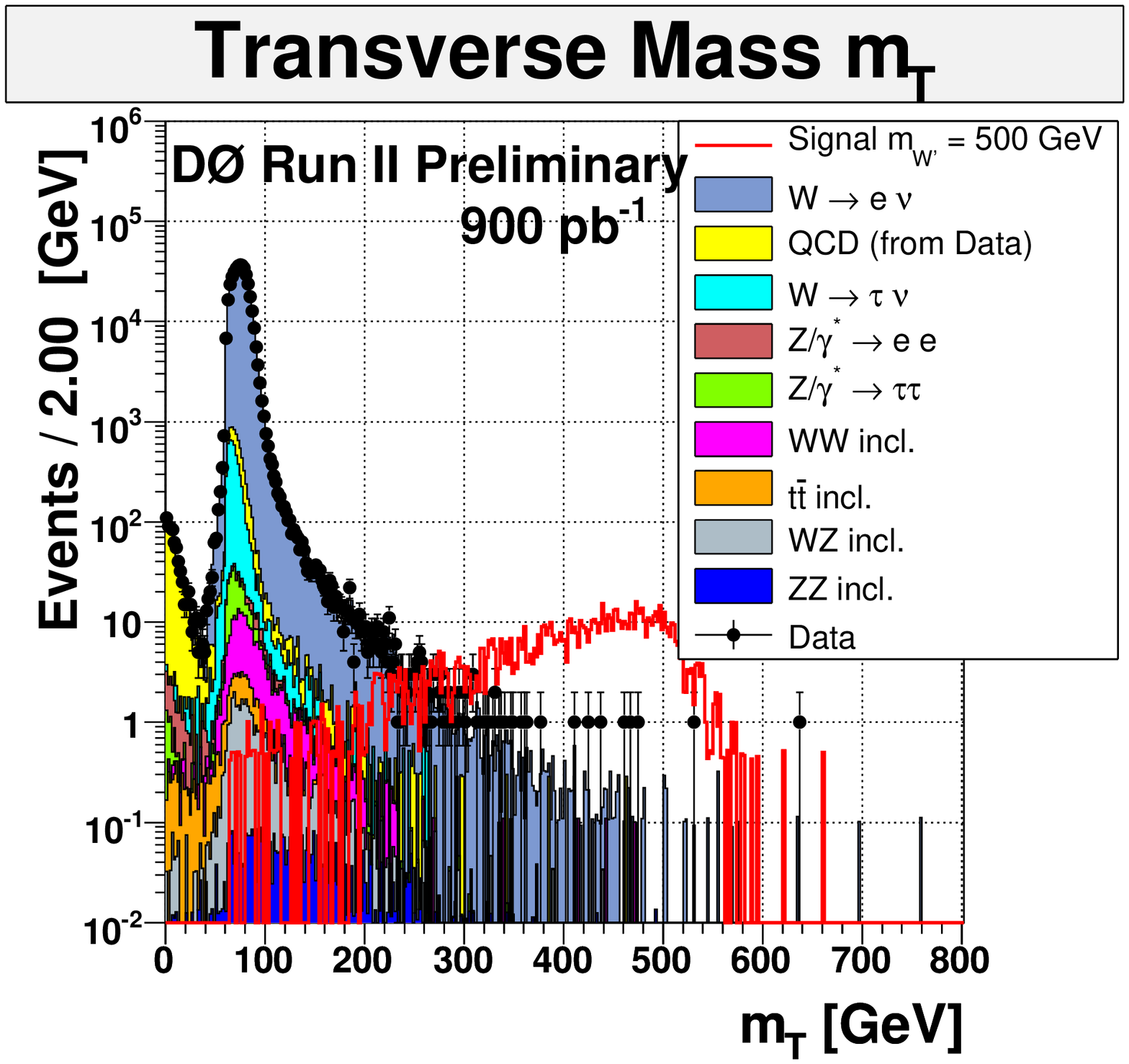,height=6.5cm}}
  \put(8.5,0){\psfig{figure=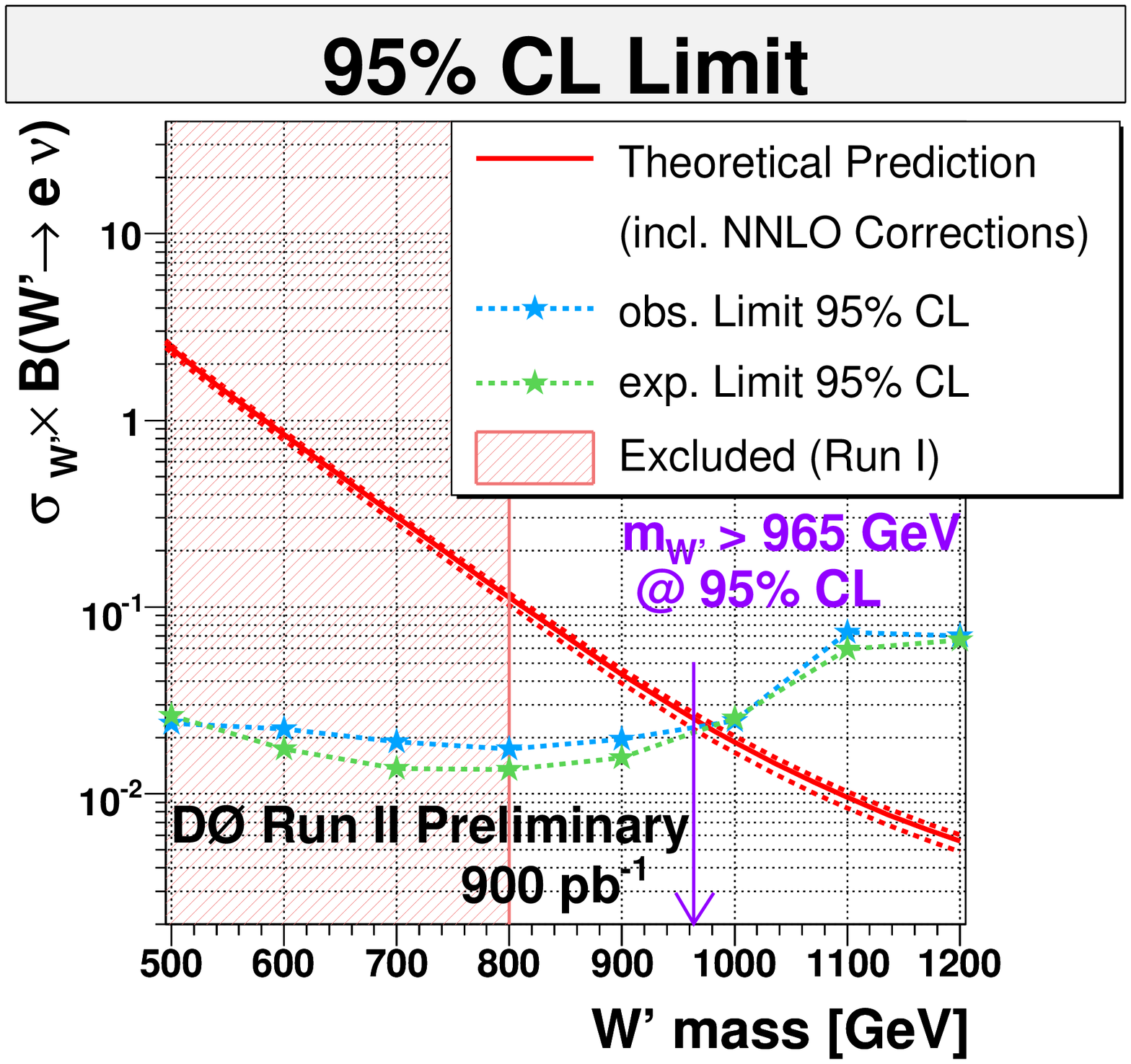,height=6.5cm}}
  \put(3.5,3.5){\bf \Large (a)}
  \put(10.5,2.5){(\bf \Large b)}
 \end{picture}
 \caption{(a) Distribution of transverse mass without $M_T$
   mass cut.  Data is shown as points with error bars, background
   is the solid histogram, while a sample signal with 
   $M_{W^\prime}$ = 500 GeV is represented by the open 
   histogram.  (b) D0 limit on the $W^\prime \rightarrow e\nu$
   cross-section times branching ratio as a function of the
   $W^\prime$ mass.
  \label{fig:wprime}}
\end{figure}

Some extensions of the standard model predict the existence
of additional, heavy, gauge bosons.  D0 has performed a search
for a $W^\prime$ decaying to an electron and 
neutrino~\cite{bib:d0wprime} using a dataset of 0.9 \fbi.  Data
selection requires a high energy electron ($E_T > 30$ GeV),
large missing transverse energy (MET $>$ 30 GeV) and large
transverse mass ($M_T > 150$ GeV).  Figure~\ref{fig:wprime}(a)
shows the transverse mass distribution (without $M_T$ cut)
for data, background and signal.  D0 observes 630 events with
an expected background of $623 \pm 18^{+83}_{-75}$ events.
Therefore, a limit on a $W^\prime$ mass $>$ 965 GeV is set
assuming standard model couplings
(Fig.~\ref{fig:wprime}(b)).

\section{$Z^\prime$}

\begin{figure}
 \unitlength1cm
 \begin{picture}(15.0,7.0)
  \put(0.5,0){\psfig{figure=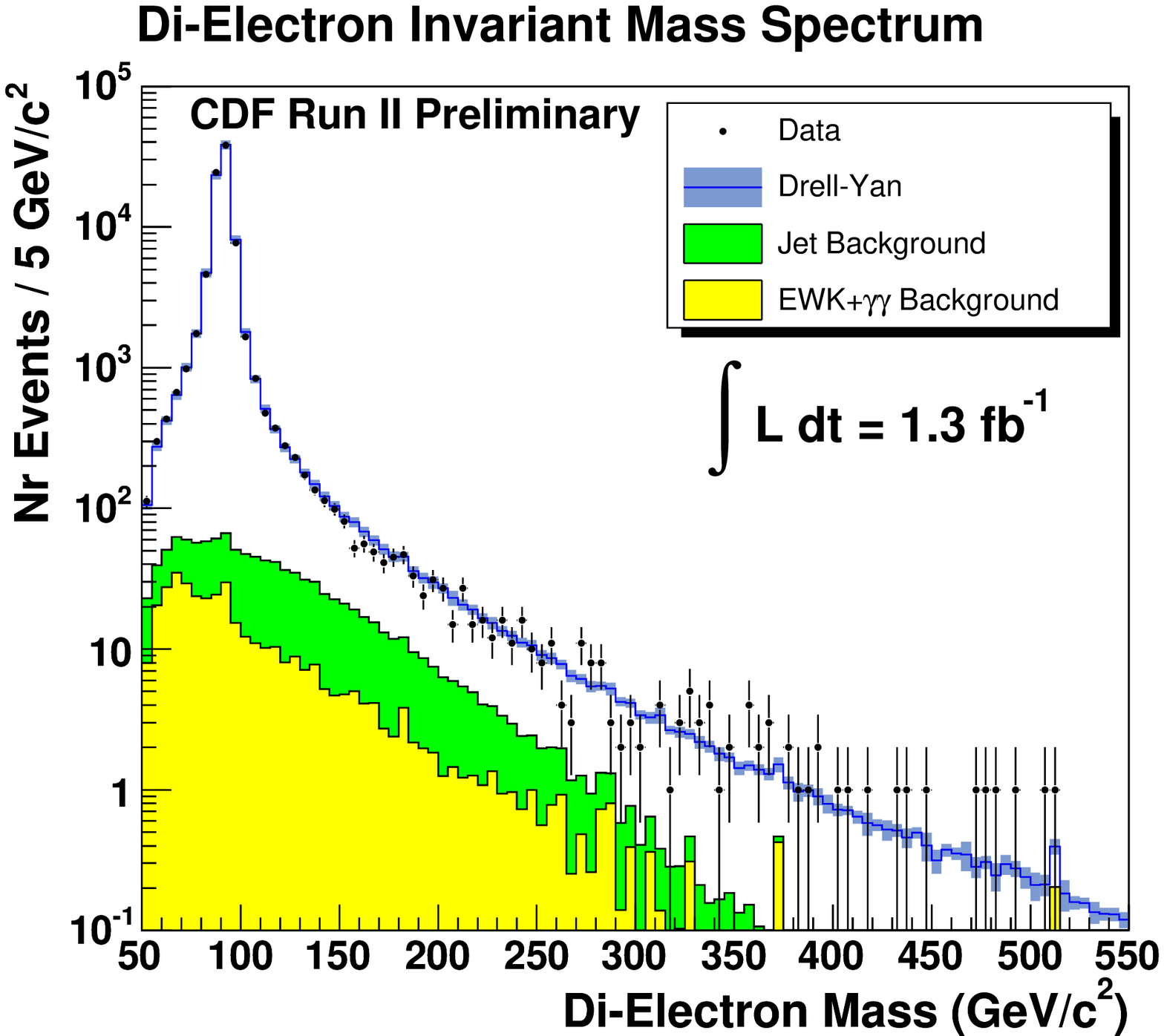,height=6.5cm}}
  \put(8.5,0){\psfig{figure=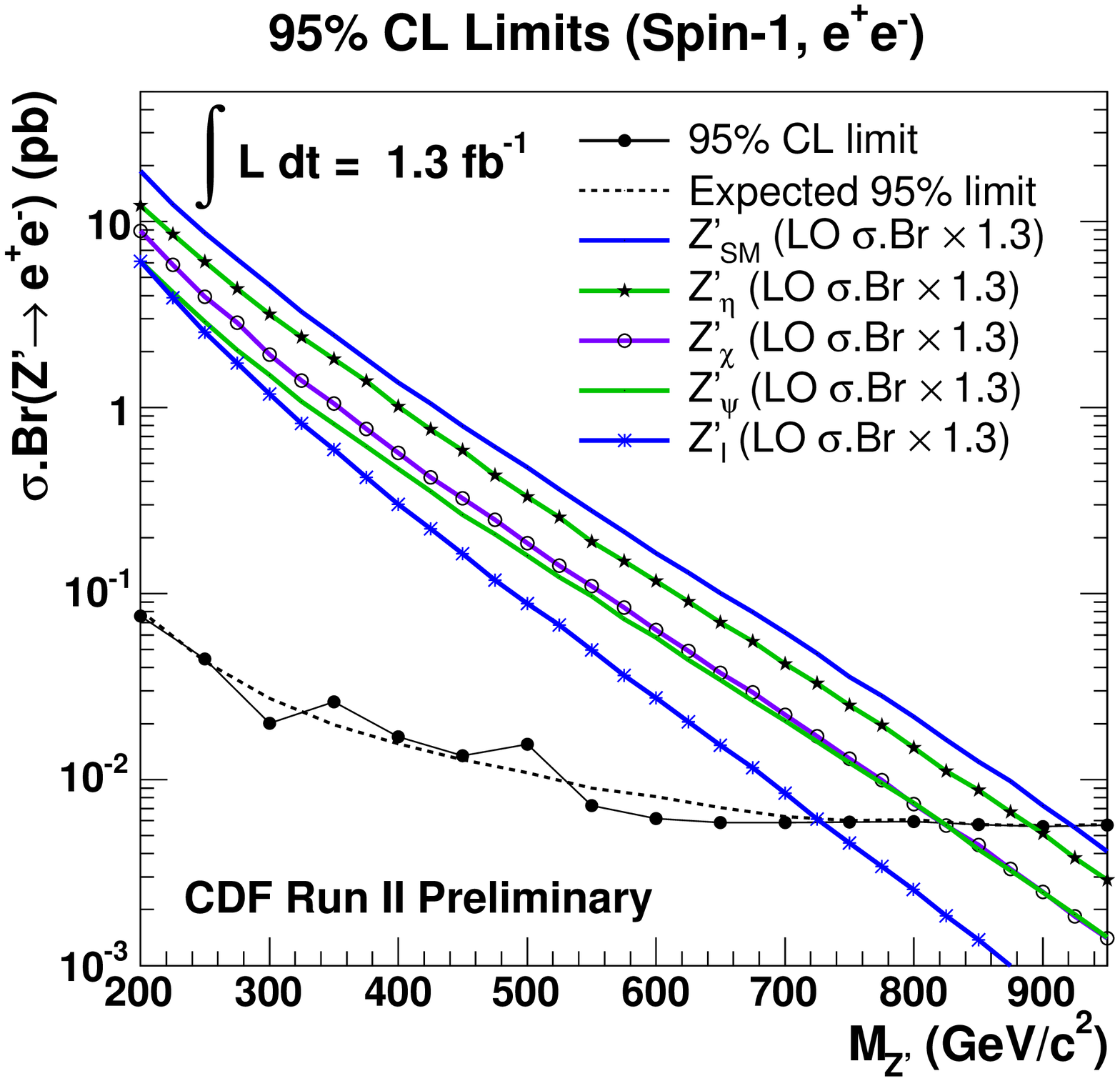,height=6.5cm}}
  \put(5.5,2.5){\bf \Large (a)}
  \put(13.5,2.5){\bf \Large (b)}
 \end{picture}
 \caption{(a) Distribution of di-electron invariant mass with
   data shown as points with errors and background as the
   histograms.  (b) CDF limit on the cross-section times
   branching ratio for a spin 1 object along with various
   models of $Z^\prime$ production.
  \label{fig:zprime}}
\end{figure}

CDF has performed a model independent search for narrow
resonances decaying to an electron and a 
positron~\cite{bib:cdfzprime} using 1.3~\fbi \ of data.  They
scan the mass region 150-900 GeV in 4 GeV mass bins looking
for an excess of data over predicted background 
(Fig~\ref{fig:zprime}(a)).  The small
excesses seen are consistent with 
statistical fluctuations.  This is interpreted to exclude
a standard model type $Z^\prime$ with mass below 
923 GeV.  Additional models are shown in Fig.~\ref{fig:zprime}(b).

\section{Randall-Sundrum Gravitons}

\begin{figure}
 \unitlength1cm
 \begin{picture}(15.0,7.0)
  \put(0.5,0){\psfig{figure=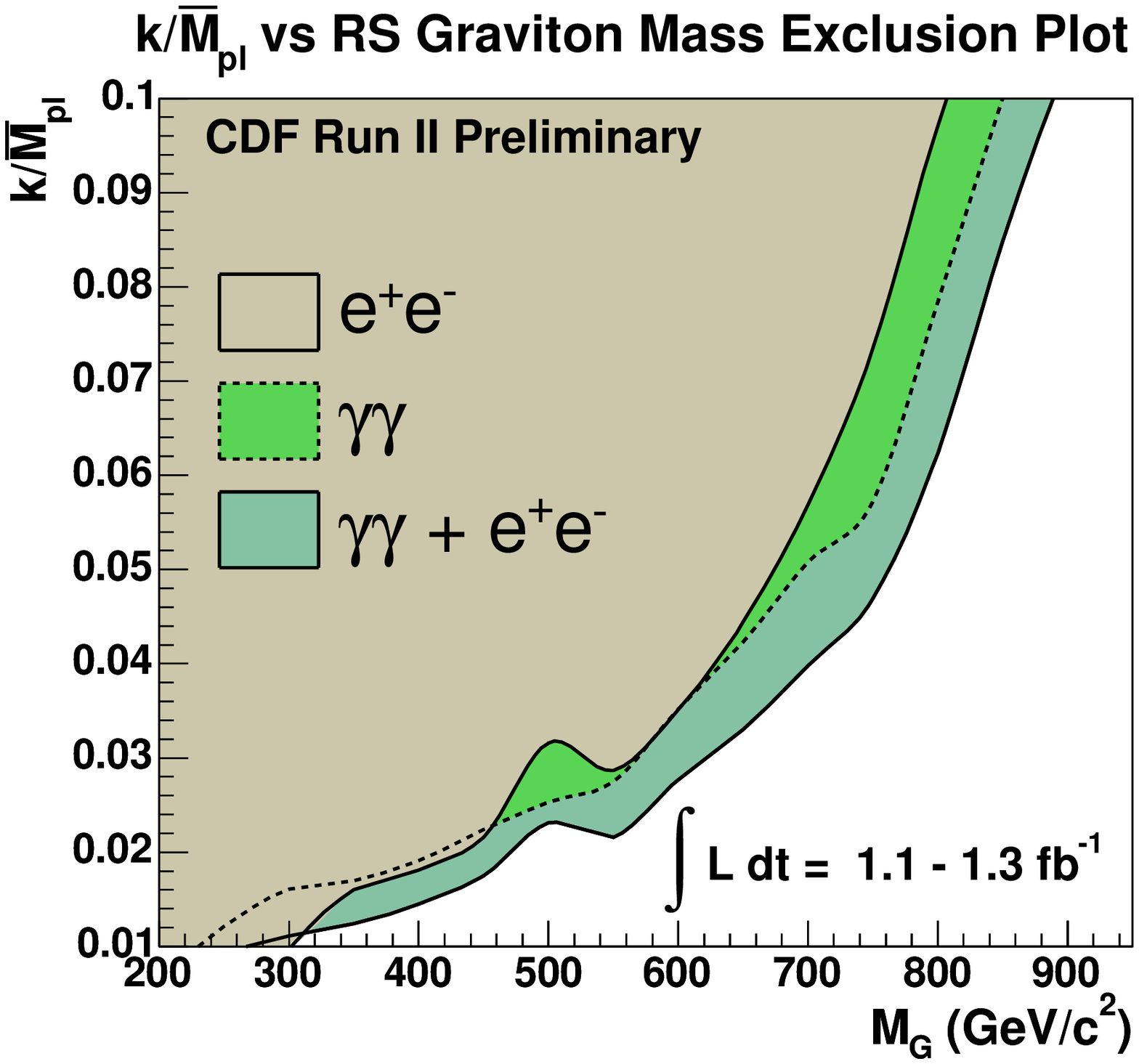,height=6.5cm}}
  \put(8.9,0){\psfig{figure=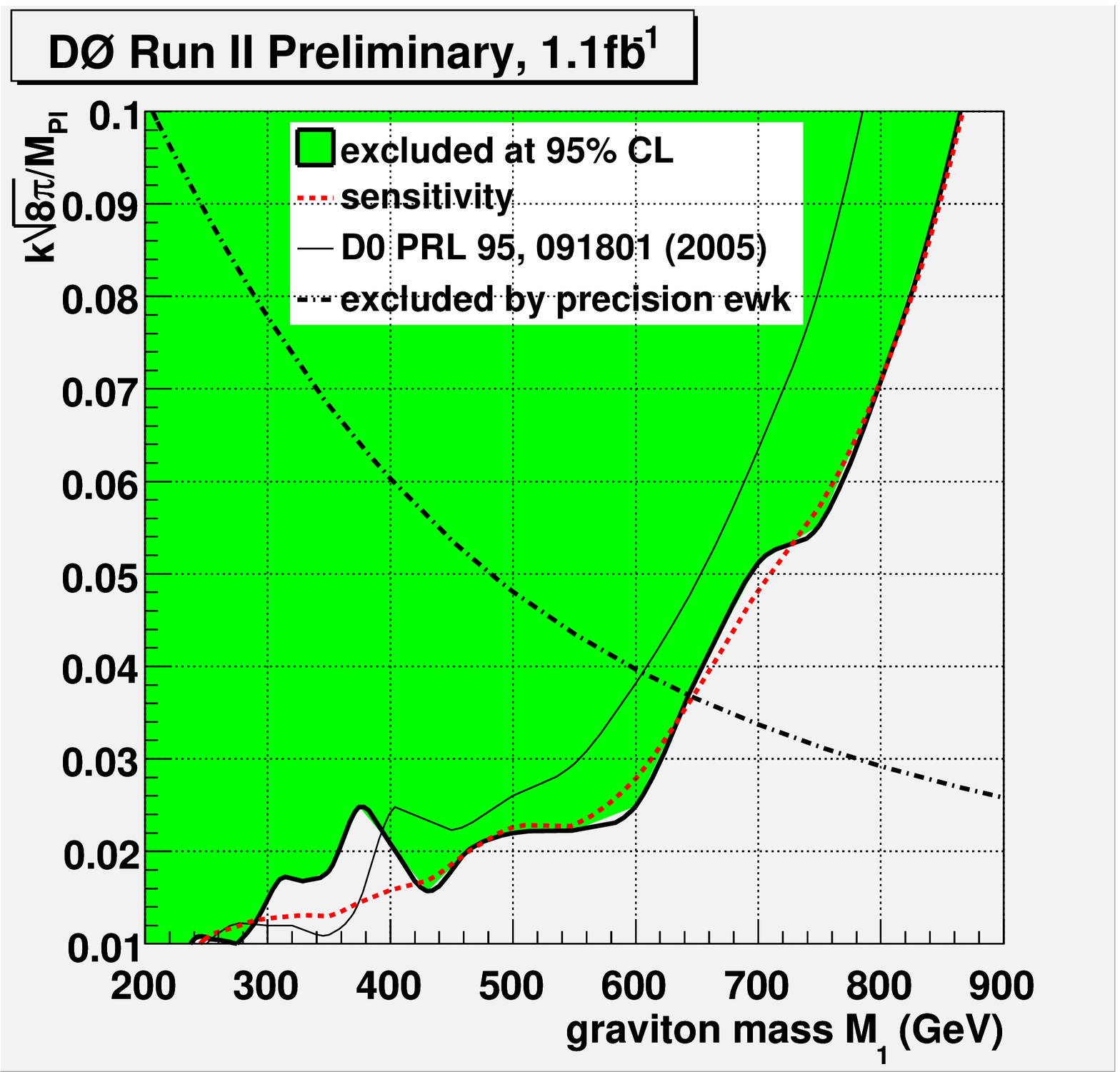,height=6.5cm}}
  \put(6,2.0){\bf \Large (a)}
  \put(13.5,1.5){\bf \Large (b)}
 \end{picture}
 \caption{Limits on extra dimensions using the Randall-Sundrum
   model from (a) CDF and (b) D0.  Limits are set on the
   parameters $k/\bar{M}_{Pl}$ and the graviton mass. 
  \label{fig:rsgravitons}}
\end{figure}

Both CDF and D0 have combined searches in di-electron
final states with similar searches in di-photon to explore
models of extra dimensions involving Randall-Sundrum
gravitons~\cite{bib:cdfzprime,bib:d0rsgrav}.  Models of
extra dimensions attempt to address the hierarchy problem 
between the strength of the weak force and gravity.
At hadron colliders, 
RS gravitons may be observed in the invariant mass or
angular distributions of electron and/or photon pairs.
Both experiments observe data in agreement with background
predictions and exclude large regions in the graviton mass
vs. k/$\bar{M}_{pl}$ parameter 
space (Fig.~\ref{fig:rsgravitons}).  At 
k/$\bar{M}_{pl}$=0.1 CDF(D0) exclude gravitons with
masses below 889(865) GeV.

\section{Excited Electrons}

\begin{figure}
 \unitlength1cm
 \begin{picture}(15.0,7.0)
  \put(0.5,0){\psfig{figure=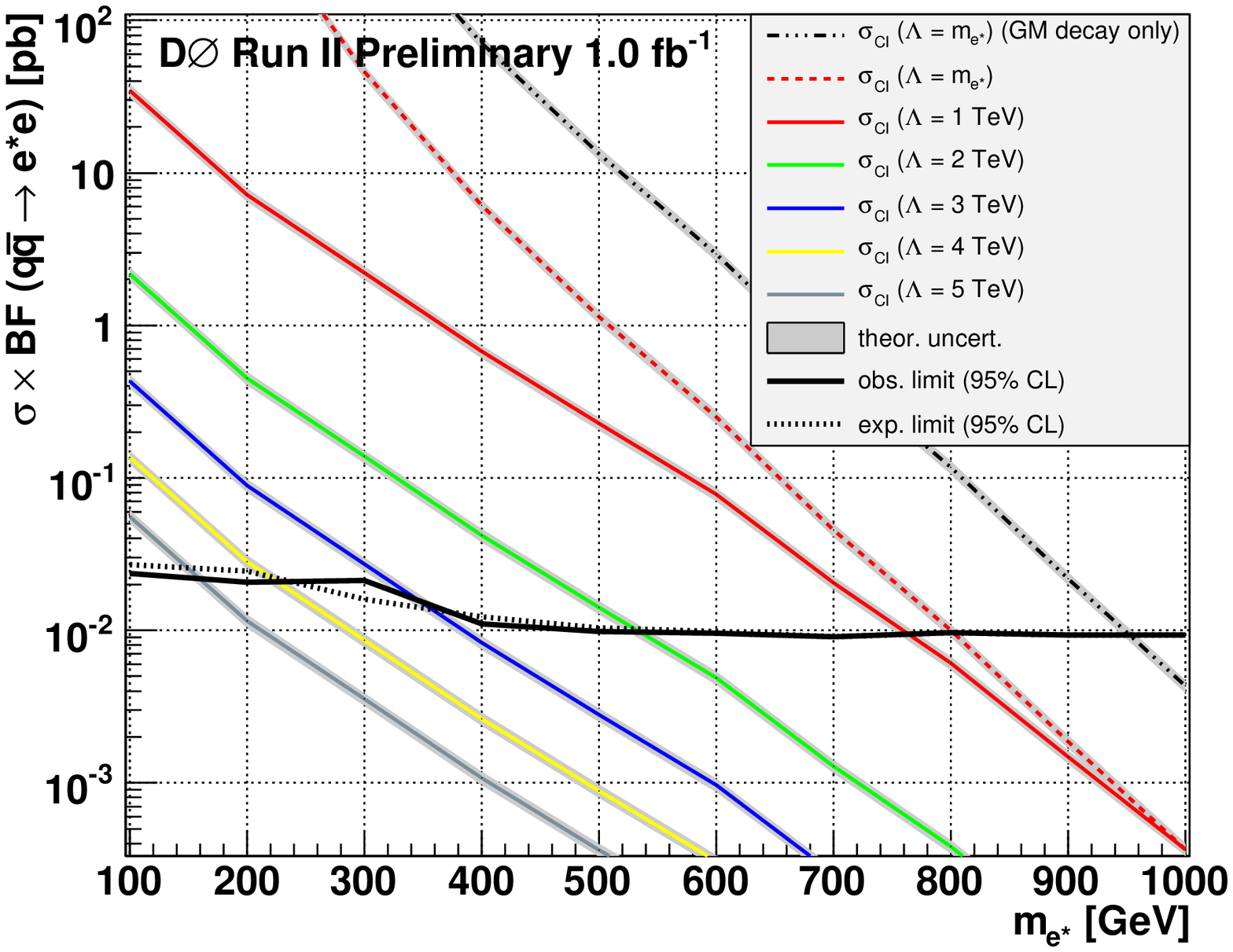,height=6.5cm}}
  \put(8.5,0){\psfig{figure=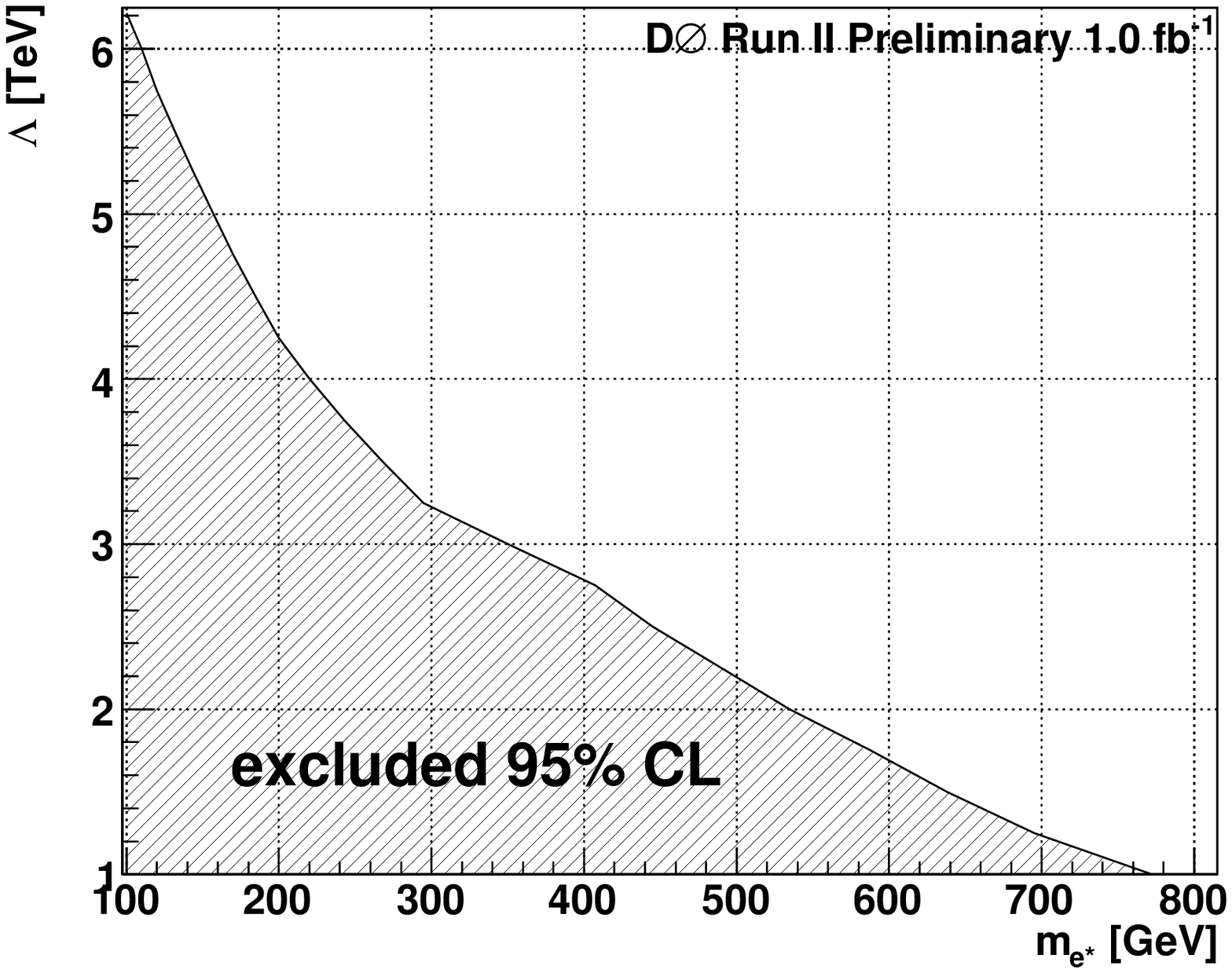,height=6.5cm}}
  \put(1.8,1.0){\bf \Large (a)}
  \put(13.5,3.5){\bf \Large (b)}
 \end{picture}
 \caption{Limits on excited electrons from D0.  (a) shows
   the limit on the cross-section times branching ratio
   as a function of the mass of the excited electron.
   (b) shows the limit on the compositeness scale vs.
   the mass of the excited electron.
  \label{fig:d0excitedelectron}}
\end{figure}

Some models predict that quarks and leptons are composite 
particles composed of smaller pieces.  These models allow
for excited quark/lepton states.  D0 has carried out a
search for excited electrons ($e^*$) from the process
$p\bar{p} \rightarrow e e^* \rightarrow e e \gamma$.  After
selecting events with $p_T(e_1,e_2,\gamma) > 25,15,15$ GeV,
259 events are observed with an expected background of
232 $\pm$ 3 $\pm$ 29 events.  From this, limits are set on
the mass of the excited electron and the compositeness
scale (Fig.~\ref{fig:d0excitedelectron}).  
For $\Lambda = 1$, the limit is $m_{e^*} > 756$ GeV.
If decays via contact interaction are neglected, D0 finds
a limit of  $m_{e^*} > 946$ GeV for $\Lambda = m_{e^*}$.

\section{Neutral, Long-lived Particles}

\begin{figure}
 \centerline{\psfig{figure=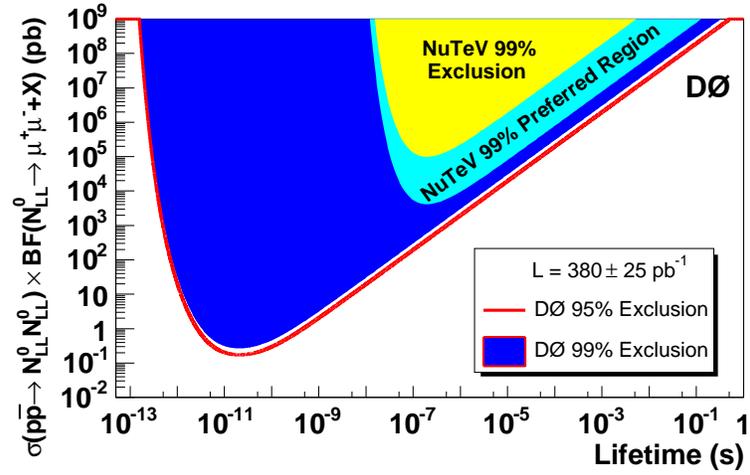,height=6.5cm}}
 \caption{Limits on the cross-section times branching
   ratio for neutral, long-lived particles decaying to
   two muons as a function of the lifetime.  The area
   above the (red) line is excluded at the 95\% CL.  The
   dark blue shaded region is a 99\% CL from D0.  The
   yellow region shows the limit from NuTeV converted to
   $p\bar{p}$ collisions at $\sqrt{s} = 1.96$ TeV.  The
   light blue region shows the area favored by a signal
   interpretation of NuTeV's result.
  \label{fig:d0nllp}}
\end{figure}

D0 has performed a search for neutral, long-lived particles
decaying to two muons after traveling at least 5 cm from
the production point.  A sample of pair production
of neutralinos with R-parity violating decays and long
lifetime is used to model the signal.  Background is 
estimated from data to be 0.75 $\pm$ 1.1 $\pm$ 1.1 events.
No events are observed with a decay length in the transverse
plane of 5-20 cm.  Limits are set on the production 
cross-section times branching ratio as well as a comparison
with a previous result from NuTeV~\cite{bib:nutev3events} 
using the sample model (Fig.~\ref{fig:d0nllp}).
This comparison limits the possible interpretations of
NuTeV's result.

\section{Summary}

The CDF and D0 collaborations have performed numerous searches
for new phenomena using leptonic final states.  Recent results
place limits on associated chargino and neutralino production,
extra gauge bosons, Randall-Sundrum gravitons, excited
electrons and neutral, long-lived particles.  Most of these
are the world's best limits.

%\begin{figure}
%\rule{5cm}{0.2mm}\hfill\rule{5cm}{0.2mm}
%\vskip 2.5cm
%\rule{5cm}{0.2mm}\hfill\rule{5cm}{0.2mm}
%%\psfig{figure=filename.ps,height=1.5in}
%\caption{Radiative (off-shell, off-page and out-to-lunch) SUSY Higglets.
%\label{fig:radish}}
%\end{figure}

\section*{Acknowledgments}
I would like to acknowledge the members of the CDF and D0 collaborations
without whom these results would not exist.  I would like to
specifically thank Christopher Hays, Yuri Gershtein,
Jean-Fran\c{c}ois Grivaz and Jane Nachtman for help in preparing the
talk and proceedings,
Finally, I thank the organizers for their efforts which resulted in such
a wonderful conference.

\end{document}